\documentclass[journal]{IEEEtran}

\usepackage{amsmath,amsfonts,amssymb,amsmath,latexsym,texdraw}

\newtheorem{theorem}{Theorem}

\begin{document}
%
\title{Weight Distributions of Hamming Codes (II)}
%
%
\author{Dae San Kim,~\IEEEmembership{Member,~IEEE}
\thanks{This work was supported by grant No. R01-2007-000-11176-0 from the Basic Research Program of the Korea Science and Engineering Foundation.}
\thanks{The author is with the Department of Mathematics, Sogang University, Seoul 121-742, Korea(e-mail: dskim@sogang.ac.kr). }
        }
\maketitle

\begin{abstract}
In a previous paper, we derived a recursive formula determining the
weight distributions of the $[n=(q^m-1)/(q-1),n-m,3]$ Hamming code
$H(m,q)$, when $(m,q-1)=1$. Here $q$ is a prime power. We note here
that the formula actually holds for any positive integer $m$ and any
prime power $q$, without the restriction $(m,q-1)=1$.
\end{abstract}



%
\IEEEpeerreviewmaketitle

\section{Introduction}
The $q$-ary Hamming code $H(m,q)$ is an $[n=(q^m-1)/(q-1),n-m,3]$
code which is a single-error-correcting perfect code. From now on,
$q$ will indicate a prime power unless otherwise stated. Also,
assume that $m>1$.

Moisio discovered a handful of new power moments of Kloosterman sums
over $\mathbb{F}_q$, when the characteristic of $\mathbb{F}_q$ is
$2$ and $3$ (\cite{M01,M02,SV,GSV}). The idea is, via Pless power
moment identity, to connect moments of Kloosterman sums and
frequencies of weights in the binary Zetterberg code of length
$q+1$, or those in the ternary Melas code of length $q-1$.

In \cite{DSK}, we adopted his idea of utilizing Pless power moment
identity and exponential sum techniques so that we were able to
derive Theorem 1 below under the restriction that $(m,q-1)=1$. This
restriction was needed to assume that $H(m,q)$ is cyclic (cf.
Theorem 3). It is somewhat surprising that there has been no such
recursive formulas giving the weight distributions of the Hamming
codes in the nonbinary cases, whereas there has been one in the
binary case(cf. Theorem 2).

In this correspondence, we will give an elementary proof showing
that the restriction $(m,q-1)=1$ can be removed.

\begin{theorem}
    Let $\{C_h\}_{h=0}^{n}(n=(q^m-1)/(q-1))$ denote the weight distribution of the
    $q$-ary Hamming code $H(m,~q)$. Then, for $h$ with $1\leq h \leq
    n$,
    \begin{align}\label{mt}
        \begin{split}
        h!C_h=&(-1)^hq^{m(h-1)}(q^m-1)\\
              &+\sum_{i=0}^{h-1}(-1)^{h+i+1}C_i\sum_{t=i}^{h}t!S(h,t)q^{h-t}(q-1)^{t-i}(^{n-i}_{n-t}),
        \end{split}
    \end{align}
    where $S(h,t)$ denotes the Stirling number of the second kind
    defined by
    \begin{equation}\label{snsk}
        S(h,t)=\frac{1}{t!}\sum_{j=0}^{t}(-1)^{t-j}(^t_j)j^h.
    \end{equation}
\end{theorem}

\begin{theorem}[p.129 in \cite{MS}]
    Let $\{C_h\}_{h=0}^{n}(n=(2^m-1))$ denote the weight distribution of the binary Hamming
    code $H(m,2)$. Then the weight distribution satisfies the following recurrence relation:
    \begin{align*}
        &C_0=1, ~C_1=0,\\
        &(i+1)C_{i+1}+C_i+(n-i+1)C_{i-1}=(^n_i)~~(i\geq 1).
    \end{align*}
\end{theorem}

\begin{theorem}[\cite{PHB}]
    Let $n=(q^m-1)/(q-1)$, where $(m,q-1)=1$.
    Let $\gamma$ be a primitive element of $\mathbb{F}_{q^m}$.
    Then the cyclic code of length $n$ with the defining zero $\gamma^{q-1}$
    is equivalent to the $q$-ary Hamming code $H(m,q)$.
\end{theorem}

\section{Proof of Theorem 1}
We know that the formula (\ref{mt}) holds for
$(m,q-1)=1$(\cite[Theorem 1]{DSK}). By the recursive formula in
(\ref{mt}), we see that all $C_i(i=0,1,2,\cdots,~n=(q^m-1)/(q-1))$
are formally polynomials in $q$ with rational coefficients, which
depend on $m$ (cf. Corollary 2 in \cite{DSK} for the explicit
expressions of $C_i$ for $i\leq 10)$. Put $C_i=P_i(q;m)$, for
$i=0,1,2,\cdots,~n=(q^m-1)/(q-1)$. Then (\ref{mt}) can be rewritten
as
\begin{align}\label{me}
    \begin{split}
        &h!P_h(q;m)=(-1)^hq^{m(h-1)}(q^m-1)+\\
              &\sum_{i=0}^{h-1}(-1)^{h+i+1}P_i(q;m)\sum_{t=i}^{h}t!S(h,t)q^{h-t}(q-1)^{t-i}\left(^{\frac{q^m-1}{q-1}-i}_{~~~t-i}\right),
    \end{split}
\end{align}
$(1\leq h\leq n=(q^m-1)/(q-1))$.

Let $m$, $h$ be fixed positive integers. Then the LHS and the RHS of
(\ref{me}) are formally polynomials in $q$ and (\ref{me}) is valid
whenever $q$ is replaced by prime powers $p^r$ satisfying
$(m,p^r-1)=1$ and $h\leq(p^{rm}-1)/(p^r-1)$.

So it is enough to show that there are infinitely many prime powers
$p^r$ such that $(m,p^r-1)=1$, since then (\ref{me}) is really a
polynomial identity in $q$, so that the restriction of our concern
can be removed. There are three cases to be considered.

Case 1) 2 does not divide $m$.

Let $m=p_1^{e_1}p_2^{e_2}\cdots p_r^{e_r}$, where
$p_1,p_2,\cdots,p_r$ are distinct odd primes and $e_j$'s are
positive integers. Then, by Dirichlet's theorem on arithmetic
progressions, there are infinitely many prime numbers $p$ such that
$p\equiv 2$(mod $m$). For each such an $p$, $p\equiv 2$(mod $p_j$),
for $j=1,\cdots,r$. Then $p_j$ does not divide $p-1$, for all $j$,
so that all $p_j$ is relatively prime to $p-1$. So $(m,p-1)=1$, for
all such primes $p$.

Case 2) 2 is the only prime divisor of $m$.

In this case, $2^l-1(l=1,2,\cdots)$ are all relatively prime to $m$.

Case 3) 2 and some odd prime divide $m$.

Let $m=2^e m_1$, $m_1=p_1^{e_1}p_2^{e_2}\cdots p_r^{e_r}$, where
$e,e_1,\cdots,e_r,r$ are positive integers and $p_1,p_2,\cdots,p_r$
are distinct odd primes. Noting that $(2,m_1)=1$, we let
$f=ord_{m_1}2$ be the order of $2$ modulo $m_1$. Then $2^{lf}\equiv
1$(mod $m_1$), for all positive integers $l$. So $2^{lf}\equiv
1$(mod $p_j$), for all $j=1,\cdots,r$. Thus $2^{lf+1}\equiv 2$(mod
$p_j$), for all $j$, and hence $p_j$ does not divide $2^{lf+1}-1$,
for all $j$. This implies that $(m,2^{lf+1}-1)=1$, for all positive
integers $l$.$\hspace*{5.5cm}\blacksquare$

%








\end{document}